\documentclass[twocolumn]{jpsj3}
\usepackage{txfonts}
\usepackage{bm}
\usepackage{here}
\usepackage{color}

\title{Anisotropic Penetration Depths of Corner States in a Higher-Order Topological Insulator}

\author{Nobuhiro Arai$^1$ and Shuichi Murakami$^{1,2}$}
\inst{$^1$Department of Physics, Tokyo Institute of Technology, 2-12-1 Ookayama, Meguro-ku, Tokyo 152-8551, Japan \\
$^2$TIES, Tokyo Institute of Technology, 2-12-1 Ookayama, Meguro-ku, Tokyo 152-8551, Japan
} 

\abst{Higher-order topological insulators in two dimensions have states that localize at their corners, called corner states. In this paper, we reveral characteristics of the penetration depth of their corner states by using the Benalcazar-Bernevig-Hughes model. First, we show that when we change the energy of the corner states toward the end of the edge gap by adding an on-site potential to the corner site, the penetration depth along the edge diverges toward infinity while the penetration depth into the bulk remaining finite. We analytically derive the corner-state wavefunction in a form of elliptic integrals, which reproduces this anisotropic behavior of corner states. This means that corner states have two kinds of penetration depths, and they behave differently. At last, we show that hybridizations between corner states are governed by the penetration depth through interference between the corner states. It is because the corner states almost do not interfere with edge states or bulk states.}

\begin{document}
\maketitle

\section{Introduction}
Topological insulators (TIs) are nonmagnetic insulators in a nontrivial topological phase\cite{a,b,c,d,e,f}. This topological phase protected by time-reversal symmetry is robust against perturbations that do not close the band gap in bulk, and is characterized by the $Z_2$ topological invariant. Interestingly, two-dimensional and three-dimensional TIs have gapless helical bound states that carry the spin current, consisting of electrons with opposite spins counter-propagating along the edge and the surface, respectively. Therefore, TIs have been attracting much attention in spintronics.

Recently, a notion of higher-order topological insulators (HOTIs)\cite{g,h,i,j,k,l,m,n,o,p,q,r,s,s2,s3,s4,prb100} has been proposed as an analogue to TIs. While TIs in $d$ dimensions have $(d-1)$-dimensional boundary states, HOTIs in $d$ dimensions have $(d-2)$- or $(d-3)$-dimensional boundary states. For example, three-dimensional second-order topological insulators are insulating in the bulk and surfaces, but have one-dimensional gapless states at the hinges. These states are called hinge states, and are protected topologically similar to edge or surface states in TIs. On the other hand, two-dimensional second-order topological insulators and three-dimensional third-order topological insulators have electronic states that localize at the corners, called corner states. The corner states are also protected topologically, and are thought of as a manifestation of higher-order multipole moments such as quadrupole moments and octupole moments, which are higher-order analogs of a dipole moment\cite{t,u,v,w,x,y,1a,1b,1c,1d}. For example, the corner states in HOTIs protected by $C_n$-symmetry result from filling anomaly\cite{i}, which is a mismatch between the number of electrons needed for charge neutrality and that needed for the material to be an insulator. Thus, physical quantities that characterize the topology of the corner states are studied intensively.

In this paper, we discuss what determines the penetration depth of the corner states in HOTIs. For comparison, in two-dimensional TIs the penetration depth of the edge states is in inverse proportion to the energy difference between the edge state and the band edge of the bulk band\cite{f}. We expect that this relationship between the energy and the penetration depth of the states that localize at the boundaries of the material holds also for the corner states. To show whether it is the case, we use the Benalcazar-Bernevig-Hughes (BBH) model\cite{g}, a model for a two-dimensional HOTI, as an example. The BBH model has both bulk bands and edge bands, and the gap for the edge bands is narrower than that for the bulk bands. In this paper we find that when we change the energy of the corner state toward the end of the edge gap, the penetration depth of the corner state along the edge diverges toward infinity while the penetration depth into the bulk remaining finite. Thus, the corner state is anisotropic, with two penetration depths behaving differently when its energy is changed. We also analytically derive the corner-state wavefunction in a form of complicated elliptic integrals, in which the anisotropic distribution is encoded. In addition, we show that hybridizations between corner states at the neighboring corners reflect the penetration depth of the corner states through interference between them. Because the corner states have different energies from those of edge states and bulk states, the corner states almost do not interfere with the edge and bulk states, but interfere mainly with other corner states. Therefore, the penetration depth directly affects hybridization between corner states.

This paper is organized as follows. In Sec. $\bm{2}$, we study the BBH model with an on-site potential to the corner site. Then we show that as the corner-state energy changes toward the end of the edge gap, the penetration depth of the corner states along the edge diverges toward infinity, but that into the bulk does not. We also derive an analytic form of the corner-state wavefunction, which is need to calculate the penetration depths directly. In Sec. $\bm{3}$, we discuss the penetration depth of the corner states by the viewpoint of the interference between the corner states. We show that the strength of hybridization between the corner states is governed by the penetration depth of the corner states. Our conclusion is given in Sec. $\bm{4}$.
\begin{figure*}[t]
\centering
\includegraphics[height=5.6cm]{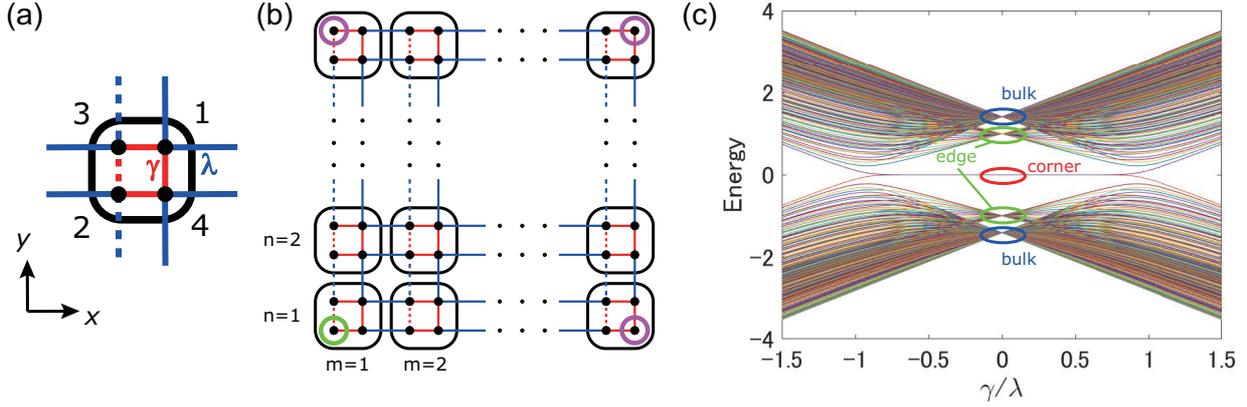}
\caption{(Color online) The BBH model and its energy bands. (a) Unit cell of the BBH model. (b) BBH model in a square shape. We will focus on the corner state at the lower left corner, by adding an on-site potential at the site specified by the green circle. (c) Energy spectrum in a square system with $\delta=0$, $\lambda=1$, when $\gamma$ is changed.}
\end{figure*}

\section{Relationship between the Penetration Depth and the Energy of the Corner State in the BBH Model}
In this section, we discuss spatial distributions of corner states. In particular, we show that the spatial decay of the corner states is not isotropic, and one can define two penetration depths, one along the edge and the other into the bulk. We show that  as we change the energy of the corner states toward the end of the edge gap, the penetration depth along the edge diverges, while that toward the bulk does not. In this sense, these two penetration depths behave differently.

\subsection{The BBH model}
First of all, we introduce the Benalcazar-Bernevig-Hughes (BBH) model showing a two-dimensional second-order topological insulator phase\cite{g}. The BBH model is a two-dimensional tight-binding model of spinless fermions consisting of four sublattices $\alpha=1,2,3,4$ per unit cell (Fig. 1(a)), with the Hamiltonain
\begin{align}
H=& \sum_{\bm{R}} \left[\delta(c^{\dagger}_{{\bm{R}},1}c_{{\bm{R}},1}+c^{\dagger}_{{\bm{R}},2}c_{{\bm{R}},2}-c^{\dagger}_{{\bm{R}},3}c_{{\bm{R}},3}-c^{\dagger}_{{\bm{R}},4}c_{{\bm{R}},4})\right.\notag\\
&+ \left.  \gamma(c^{\dagger}_{{\bm{R}},1}c_{{\bm{R}},3}+c^{\dagger}_{{\bm{R}},1}c_{{\bm{R}},4}-c^{\dagger}_{{\bm{R}},2}c_{{\bm{R}},3}+c^{\dagger}_{{\bm{R}},2}c_{{\bm{R}},4}+{\rm{H.c.}}) \right.\notag\\
&+ \left. \lambda(c^{\dagger}_{{\bm{R}},1}c_{{\bm{R}}+\hat{\bm{x}},3}+c^{\dagger}_{{\bm{R}},1}c_{{\bm{R}}+\hat{\bm{y}},4}-c^{\dagger}_{{\bm{R}},3}c_{{\bm{R}}+\hat{\bm{y}},2}\right.\notag\\
&\ \left. \ \ \ \  \ \ \ \ \ \ \ \ \ \ \ \ \ \ \ \ \ \ \ \ \ +c^{\dagger}_{{\bm{R}},4}c_{{\bm{R}}+\hat{\bm{x}},2}+{\rm{H.c.}}) \right],\label{1}
\end{align}
where $c^{\dagger}_{{\bm{R}},\alpha} (\alpha=1,2,3,4)$ is the creation operator at the $\alpha$-sublattice in the unit cell $\bm{R}$, as shown in Fig. 1(a), (b). Here ${\bm{R}}=(m,n)\ (m,n:{\rm{integer}})$ specifies the position of the unit cell forming a square lattice and $\hat{\bm{x}}$ and $\hat{\bm{y}}$ are unit vectors along the $x$ and $y$ directions, respectively. The parameter $\delta$ represents an on-site potential, and $\gamma$ and $\lambda$ are hopping parameters within a unit cell and between the nearest-neighbor unit cells, respectively. Figure 1(c) shows an energy spectrum of the system in a square shape in Fig. 1(b). As this figure indicates, energy bands of corner states, edge states, and bulk states are well separated near $\gamma \sim 0$.

\subsection{Bulk bands}
In the momentum space, the bulk Hamiltonian is
\begin{align}
H({\bm{k}},\delta)&=(\gamma+\lambda \cos{k_x})\Gamma_4+\lambda \sin{k_x}\Gamma_3 \notag\\
&+ \ \ (\gamma+\lambda \cos{k_y})\Gamma_2+\lambda \sin{k_y}\Gamma_1+\delta \Gamma_0\label{2}
\end{align}
where $\Gamma_0=\tau_3 \sigma_0$, $\Gamma_k=-\tau_2 \sigma_k$$(k=1,2,3)$, $\Gamma_4=\tau_1\sigma_0$, and $\tau_i$ and $\sigma_i\ (i=1,2,3)$ are Pauli matrices. In particular, $H({\bm{k}},\delta=0)$ has energies $E=\pm \sqrt{2\lambda^2+2\gamma^2+2\gamma \lambda (\cos{k_x}+\cos{k_y})}$, each of which is two-fold degenerate. When $\delta=0$ and $|\gamma/\lambda|<1$, this model is a two-dimensional second-order topological insulator and has corner states with zero energy that decay exponentially\cite{g,h,s2}. For simplicity, we henceforth consider the case with $0<\gamma<\lambda$. In this case, the bulk bands lie in the regions $\sqrt{2}(\lambda-\gamma)<|E|<\sqrt{2}(\lambda+\gamma)$. We set $\delta=0$ in Sec. $\bm{2}$ and $\delta \neq0$ in  Sec. $\bm{3}$.
\begin{figure}[H]
\centering
\includegraphics[height=3.8cm]{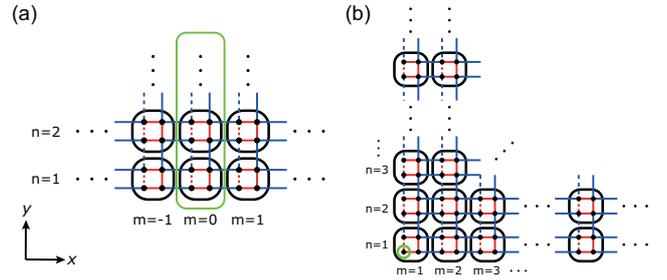}
\caption{(Color online) Two geometries of the BBH model. (a) A semi-infinite geometry, which is infinite along the $x$ direction and semi-infinite along the $y$ direction. Because of translation symmetry along the $x$ direction, the wavefunction is characterized by a Bloch wavenumber $k$ along $x$ direction with the unit cell specified in green. (b) A wedge geometry. We will focus on the corner state, by adding an on-site potential at the site specified by the green circle.}
\end{figure}
\begin{figure*}[t]
\centering
\includegraphics[width=17.0cm]{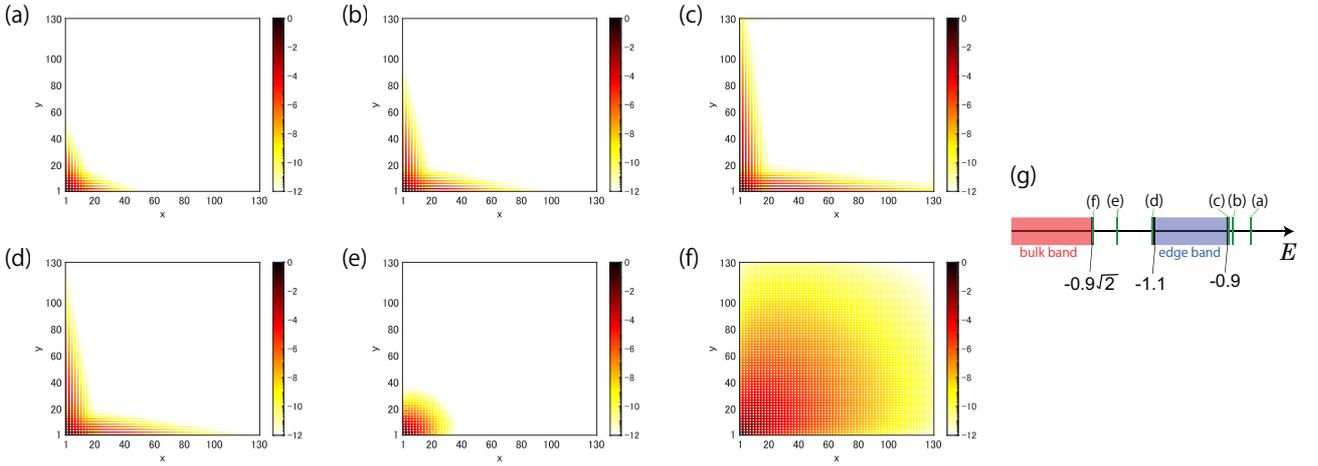}
\caption{(Color online) Distributions of the corner state, with their amplitude shown in colors in a common logarithm. The system contains $65$ unit cells both along $x$ and $y$ direction, and adopt the parameter $\gamma=0.1$, $\lambda=1$, $\delta=0$, and $\Delta'=10$. (a) $\Delta=-0.9$, $E=-0.8353$, (b) $\Delta=-1$, $E=-0.8837$, (c) $\Delta=-1.05$, $E=-0.8951$, (d) $\Delta=-0.98$, $E=-1.1079$,
(e) $\Delta=-1.15$, $E=-1.2043$, and
(f) $\Delta=-1.23$, $E=-1.2707$. (g) Corner-state energies in (a)-(f) are shown, together with the energies of the edge band and the bulk band. The corner-state energy is inside the gap between the two edge bands in (a)-(c), and
inside the gap between the edge band and the bulk band in (d)-(f).}
\end{figure*}

\subsection{Edge states}
In this subsection, we consider a semi-infinite system, which is infinite along the $x$ direction and semi-infinite along the $y$ direction. It has sites at $(m,n)$, with an integer $m$ and a positive integer $n$ as shown in Fig. 2(a). We calculate edge states in this semi-infinite geometry. Through straightforward calculations, whose details are explained in Appendix A, we get the edge-state wavefunctions ${\bm{u}}_{m,n}$ and their energies $E$:
\begin{align}
{\bm{u}}_{m,n}&={\bm{u}}_n(k)e^{imk},\label{3}\\
{\bm{u}}_n(k)&=\left(
\begin{array}{c}
u^{(1)}_n(k)\\
u^{(2)}_n(k)\\
u^{(3)}_n(k)\\
u^{(4)}_n(k)
\end{array}
\right)
=\left({-\frac{\gamma}{\lambda}}\right)^n\left(
\begin{array}{c}
0\\
E\\
0\\
\gamma+\lambda e^{ik}
\end{array}
\right)_,(n\geq1)\label{4}
\end{align}
and
\begin{align}
E=\pm \sqrt{\lambda^2+\gamma^2+2\gamma \lambda \cos{k}},\label{5}
\end{align}
where $k$ is a Bloch wavevector along the $x$ axis. Because we are considering the case with $0<\gamma<\lambda$, the bands of the edge states lie in the regions $-\lambda-\gamma \leq E \leq -\lambda+\gamma$, $\lambda-\gamma \leq E \leq \lambda+\gamma$. We note that the form of the edge-state wavefunctions are given by a single exponential term with a decay rate $-\gamma/\lambda$.

\subsection{Corner states}
In this subsection, we calculate wavefunctions and energies of corner states of the BBH model. We consider a wedge geometry with sites at $(m,n)$ $(m\geq1,n\geq1)$ shown in Fig. 2(b). Latter, we will add an on-site potential $\Delta$ at the corner site, but for the present we put $\Delta=0$. Then, in Ref. \citen{s2}, the corner-state wavefunction is obtained as
\begin{align}
{\bm{u}}_{m,n}=\left(
-\frac{\gamma}{\lambda}\right)^{m+n}
\left(
\begin{array}{c}
0\\
1\\
0\\
0
\end{array}
\right)_,\label{6}
\end{align}
and its energy is $E=0$. This wave function describes a spatial decay with a single decay rate $-\gamma/\lambda$ both along $x$ and $y$ directions. When we write the wavefunction as ${\bm{u}}_{m,n}\propto e^{-\frac{m}{l}}e^{-\frac{n}{l}}$ with $l$ being a penetration depth, we get $1/l=\ln{|\lambda/\gamma|}$.

Nonetheless, as we show in the following, this simple form Eq. (\ref{6}) of the corner state is limited to a very special case. In this paper, we discuss general behaviors of this penetration depth. To this end, here we review behaviors of the penetration depth of topological edge states in a two-dimensional topological insulator. As the energy of the edge state approaches the bulk bands, the penetration depth is shown to diverge \cite{f}. This behavior is natural because the edge state gradually becomes bulk states. We can expect a similar behavior for the corner states in the two-dimensional higher-order topological insulators. Here we note that apart from corner states, we also have edge bands and bulk bands, with their gaps given by $-(\lambda-\gamma)\leq E \leq \lambda-\gamma$ and $-\sqrt{2}(\lambda-\gamma)\leq E \leq \sqrt{2}(\lambda-\gamma)$, respectively. Thus, when the energy of the corner state changes and approaches the edge band, it is still far from the bulk band. Correspondingly, we can expect that the penetration depth of the corner states becomes highly anisotropic. Namely, the penetration depth along the edge will diverge but that into the bulk stays finite. In this paper, we show that it is indeed the case.

In order to show this, we need to change the corner-state energy. One might think that can be achieved by a change of $\delta$ in Eq. (\ref{1}). Nonetheless, it is not sufficient for our purpose, because when $\delta$ is changed away from zero, the edge gap and bulk gap become larger, and the corner-state energy never reaches the edge bands. Instead, we add an on-site potential $\Delta$ only to one site at the lower left corner, i.e. the sublattice $\alpha=2$ in the cell $(1,1)$ marked with a green circle shown in Fig. 1(b), and see how the spatial distribution of the corner state changes. 
\begin{figure*}[t]
\includegraphics[height=6cm]{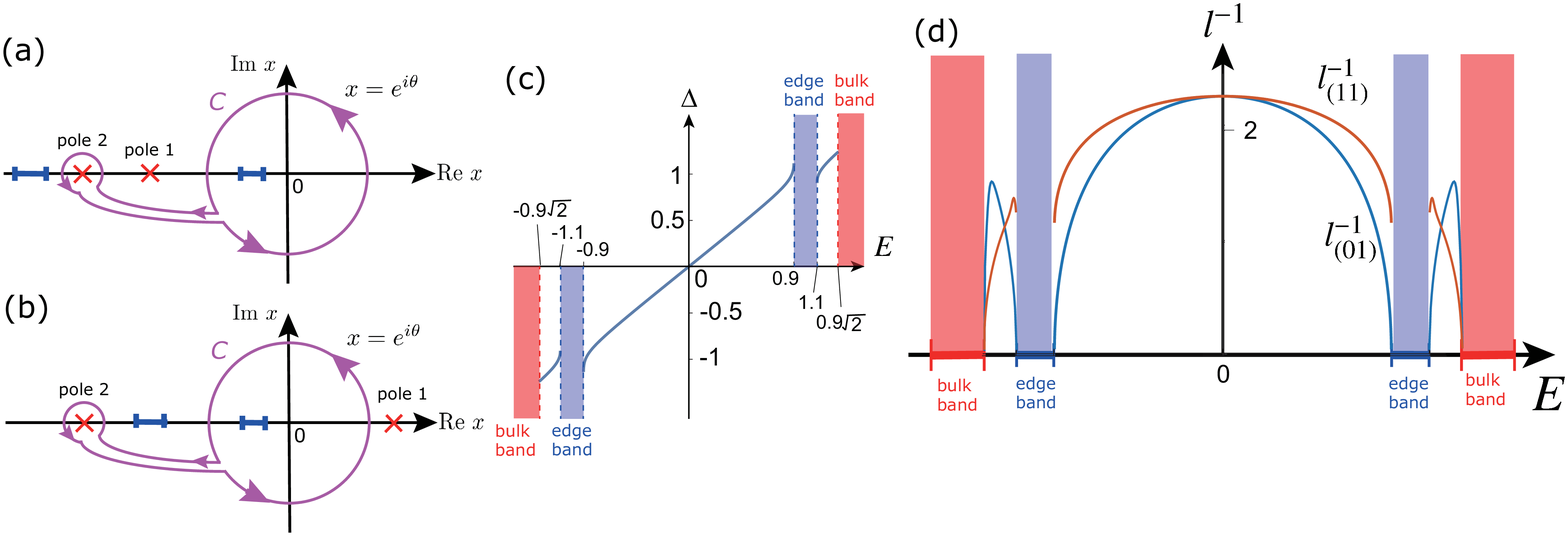}
\caption{(Color online) An analytical solution of the corner state, depending on the on-site potential $\Delta$ at the corner site. In this calculation we put $\lambda=1$, $\gamma=0.1$ and $\delta=0$. (a)(b) Contours of the integral in Eq. (\ref{7}) for 
(a) $|E|<\lambda-\gamma$ and (b) $\lambda+\gamma<|E|<\sqrt{2}(\lambda-\gamma)$. Blue lines represent branch cuts for $\tau=\sqrt{\sigma^2-4}$, where $\sigma=\frac{E^2-2(\lambda^2+\gamma^2)}{\lambda\gamma}-x-\frac{1}{x}$. Namely, their end points are given by $\sigma=\pm2$. The pole 1 and 2 are given by $x+\frac{1}{x}=\frac{E^2-\lambda^2-\gamma^2}{\lambda \gamma}$ ($|x|>1$) and $x=-\frac{\lambda}{\gamma}$, respectively. (c) The relationship between $\Delta$ and the corner-state energy $E$. (d) The inverse of the penetration depths, $l^{-1}_{(01)}$ and $l^{-1}_{(11)}$. Blue and red regions represent the edge bands and the bulk bands, respectively. Here, within $-0.9<|E|<0.9$ the penetration depths are calculated by $l^{-1}_{(01)}\equiv {\ln{\left|{u^{(2)}_{1,n}/u^{(2)}_{1,n+1}}\right|}}$ with $n=25$ and $l^{-1}_{(11)}\equiv \frac{1}{2}{\ln{\left|{u^{(2)}_{m,m}/u^{(2)}_{m+1,m+1}}\right|}}$ with $m=20$. In $1.1<|E|<0.9\sqrt{2}$, some of the elements $u_{m,n}$
change across zero, and we use an alternative definition of the penetration depths as $l^{-1}_{(01)}\equiv \frac{1}{2}{\ln{\left|{u^{(2)}_{1,n}/u^{(2)}_{1,n+2}}\right|}}$ with $n=7$ and $l^{-1}_{(11)}\equiv \frac{1}{4}{\ln{\left|{u^{(2)}_{m,m}/u^{(2)}_{m+2,m+2}}\right|}}$ with $m=8$.
}
\end{figure*}

We first show behaviors of corner states in a numerical diagonalization of the system in a sufficiently large square shape. To minimize an influence of interference between corner states, we also add a large on-site potential $\Delta'$ to the other three corners, at the sites marked with purple circles in Fig. 1(b). We set the parameters $\gamma=0.1$, $\lambda=1$, $\Delta'=10$ and change the on-site potential $\Delta$ to control the energy of the corner state that localizes at the lower left corner.
We show spatial distribution of the wavefunction of the corner state for various values of the corner-state energy. 
Here the edge bands are at $0.9\leq|E|\leq1.1$, and the bulk bands are at $0.9\sqrt{2}\leq|E|\leq1.1\sqrt{2}$.
We first show the cases when the corner state is between the two edge bands, i.e. $-0.9<|E|<0.9$,
in Fig. 3 (a) ($\Delta=-0.9$), (b) ($\Delta=-1$) and (c) ($\Delta=-1.05$) , giving the energies of the corner states to be (a) $E=-0.8353$, (b) $E=-0.8837$, and (c) $E=-0.8951$, respectively. From the figures we observe that the penetration depth of the corner state along the edge diverges toward infinity but that into the bulk does not, as the energy approaches the end of the edge band at $E=-0.9$.
Next, we also calculate the cases with the corner state being in the region $1.1<|E|<0.9\sqrt{2}$, between the edge band and the bulk band. The results are shown in Fig. 3(d)-(f) for
(d) $\Delta=-0.98$, $E=-1.1079$,
(e) $\Delta=-1.15$, $E=-1.2043$, and
(f) $\Delta=-1.23$, $E=-1.2707$.
In Fig.3(d), where the energy is close to the edge band, the corner state spreads mainly along the edge. As the energy becomes lower, 
the corner state gradually becomes localized (Fig. 3(e)). By making the energy lower toward the bulk band, the corner state 
is distributed more broadly, as seen in Fig. 3(f), meaning that it deeply penetrates into the bulk. We show the corner-state energies in (a)-(f) in Fig. 3(g).

\begin{figure*}[t]
\centering
\includegraphics[height=14.0cm,width=18.0cm]{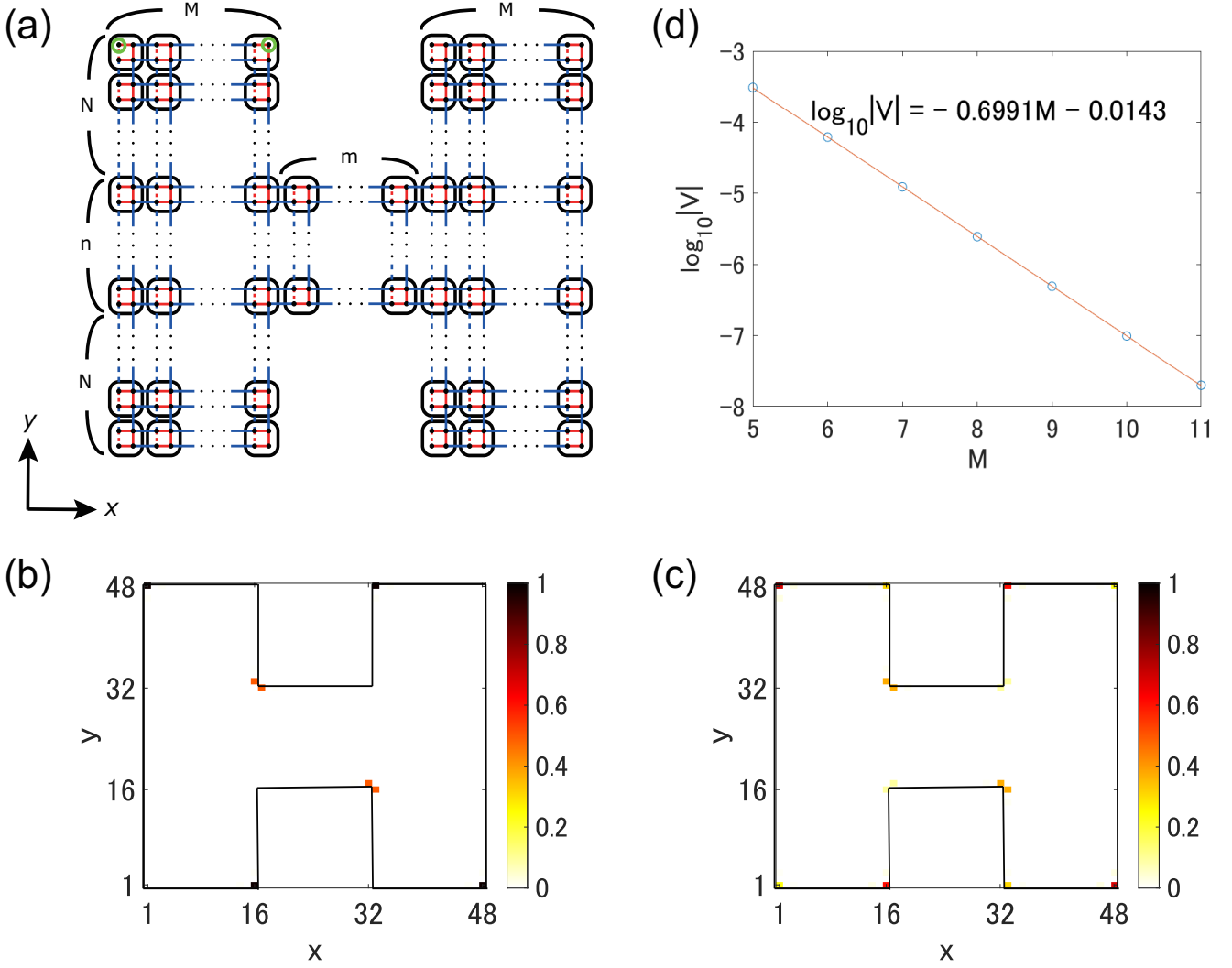}
\caption{(Color online) The interference between the corner states. (a) An `H'-shaped system. (b), (c) Electron density of the corner states below $E=0$ for (b) $\gamma=0.1$ and (c) $\gamma=0.2$. We adopt $\lambda=1$, $\delta=10^{-6}$, and $m=M=n=N=8$. $x$ and $y$ represent the coordinate of the site. (d) Amplitude of the hopping $V$ between the corner states. $M$ is the number of cells between the corners marked with  green circles.}
\end{figure*}

In the following, we analytically calculate the wavefunction of the corner state that localizes at the lower left corner. 
Here we consider a corner state in the gaps, i.e. $|E|<\lambda-\gamma$ or $\lambda+\gamma<|E|<\sqrt{2}(\lambda-\gamma)$.
Let ${\bm{u}}_{m,n}$ denote the wavefunction in the $(m,n)$ cell $(m\geq1,n\geq1)$. Let $u^{(\alpha)}_{m,n}$ denote the $\alpha$-th component of ${\bm{u}}_{m,n}$ where $\alpha=1,2,3,4$. The solution is given by the following 
\begin{align}
{\bm{u}}_{m,n}&=\frac{1}{2\pi i}\oint_{C}\ dx\frac{x-\frac{1}{x}}{\left({\gamma+\lambda \frac{\sigma+\tau}{2}}\right)\left({\gamma+\frac{\lambda}{x}}\right)}
\left(
\begin{array}{c}
0\\
E\\
-\gamma-\lambda \frac{\sigma+\tau}{2}\\
\gamma+\frac{\lambda}{x}
\end{array}
\right)\notag\\
&\ \ \ \ \ \ \ \ \ \ \ \ \ \ \ \ \ \ \ \ \ \ \ \ \ \ \ \ \ \ \ \ \ \  \cdot \frac{1}{x^{m+1}}\left({\frac{\sigma+\tau}{2}}\right)^n_,\label{7}
\end{align}
where $E$ is the energy of the corner state, and $\sigma$, $\tau$ are given by
\begin{align}
\sigma&=\frac{E^2-2(\gamma^2+\lambda^2)}{\gamma \lambda}-x-{\frac{1}{x}}_.\label{8}\\
\tau&=\sqrt{\sigma^2-4}.\label{9}
\end{align}
Here the integral contour $C$ is shown in Fig. 4(a) and (b) for 
(a) $|E|<\lambda-\gamma$ and (b) $\lambda+\gamma<|E|<\sqrt{2}(\lambda-\gamma)$, respectively. It
consists of a contour integral along the unit circle and one encircling the pole 2 at $x=-\lambda/\gamma$. 
The branch of the square root in Eq.~(\ref{9}) is taken such that it is positive along the unit circle. Consequently, the branch of
the square root in Eq.~(\ref{9}) near the pole 2 is taken in such a way that $\tau>0$ in 
(a) $|E|<\lambda-\gamma$, and $\tau<0$ in (b) $\lambda+\gamma<|E|<\sqrt{2}(\lambda-\gamma)$, 
The details of the calculation is presented in Appendix B.

The on-site potential $\Delta$ is given by
\begin{align}
\Delta={\frac{\lambda}{\gamma u^{(2)}_{1,1}}}_.\label{10}
\end{align}
We explain the detailed derivation of the above equations in Appendix B. In Fig. 4(c), we show the relationship between the on-site potential $\Delta$ and the corner-state energy $E$ obtained from Eq. (\ref{10}). As the on-site potential $\Delta$ is lowered, the corner-state energy is changed. In some regions of the value of $\Delta$, there can be two corner states with different energies. 

We calculate the change of the penetration depth of the corner state when we change its energy. Here, we define two penetration depths of the corner state. One is a penetration depth along the edge parallel to the $(0\ 1)$ direction defined by $l^{-1}_{(01)}\equiv {\ln{\left|{u^{(2)}_{1,n}/u^{(2)}_{1,n+1}}\right|}}$, and the other is along the $(1\ 1)$ direction into the bulk defined by $l^{-1}_{(11)}\equiv \frac{1}{2}{\ln{\left|{u^{(2)}_{m,m}/u^{(2)}_{m+1,m+1}}\right|}}$. This factor $\frac{1}{2}$ is inserted so that in the case of an isotropically decaying case represented by $u^{(\alpha)}_{m,n}\sim r^{-m-n}$, these two penetration depths become equal. Our result from Eq. (\ref{7}) is shown in Fig. 4(d). This indicates that as the corner-state energy approaches the edges of the edge band at $E=\pm0.9$ and $E=\pm1.1$ from inside the gaps, the penetration depth $l_{(01)}$ along the edge diverges toward infinity but that toward the bulk $l_{(11)}$ does not. Through this evolution of the corner state, it eventually becomes an edge state, and is absorbed into the edge bands, with its spatial distribution approaching that of an edge state. A similar phenomenon was seen also in a different model \cite{prb100}. On the other hand, when the energy is in the region $1.1<|E|<0.9\sqrt{2}$ and approaches the bulk band , the two penetration depths diverge, which
means that the corner state becomes a bulk state. This is in accordance with the result shown in Fig. 3(f), where the state
gradually penetrates into the bulk. 
We note that in the particular case of $E=0$ the corner state is given by Eq. (\ref{6}) \cite{s2}, which means that $l^{-1}_{(11)}=l^{-1}_{(01)}=\ln{|\lambda/\gamma|}$. From this result, we conclude that the corner states have the two penetration depths, one along the edge and the other toward the bulk, and they behave differently. 
In particular, the corner state becomes highly anisotropic as its energy approaches the edge band. Such an anisotropic behavior cannot be represented by a single exponential form like $r^{-m-n}$, but it is encoded in the complicated form (Eq. (\ref{7})) involving elliptic integrals, as discussed in Appendix B.

\section{Interference between Corner States}
In this section, we demonstrate that the penetration depth of corner states can be known from interference between the corner states.
In the following, we set $\delta=10^{-6}$ to slightly break the $C_4$ symmetry to illustrate interference between corner states. We consider an `H'-shaped system shown in Fig. 5(a) to study interference between various corner states that localize at 90$^\circ$-degree corners and 270$^\circ$-degree corners. Our results for the electron density of corner states below $E=0$ are in Fig. 5(b) and (c). In these figures, values of the parameter $\gamma / \lambda$ are different, leading to different electron densities of corner states. In Fig. 5(b) with $\gamma /\lambda=0.1$, some corners have an appreciable electron density, but the others do not. On the other hand, in Fig. 5(c) with $\gamma / \lambda=0.2$, all the corners have a comparable electron density. This is because the corner states decay exponentially along the edges, and the wavefunctions of corner states begin to overlap for a larger value of $\gamma / \lambda$ in Fig. 5(c). Therefore, it leads interference between the corner states, leading to the electron density mentioned above.

As we mentioned, the energies of corner states, edge states, and bulk states are well separated near $\gamma/\lambda \sim 0$. This means that the corner states almost do not interfere with edge states and bulk states, but with other corner states only. Therefore, we expect that an effective hopping between corner states is determined by interference between corner states. On the basis of this expectation, we describe the interference between two corner states at the neighboring corners with on-site energies $-\delta$ and $\delta$ by the following matrix
\begin{align}
H=\left(
\begin{array}{cc}
-\delta & V \\
V^* & \delta
\end{array}
\right)_,\label{11}
\end{align}
where $V$ is a hopping between two corner states. Its eigenstate $\psi$ with a lower energy is also represented as 
\begin{align}
\psi =\left(
\begin{array}{c}
\psi_1\\
\psi_2
\end{array}
\right)
\propto \left(
\begin{array}{c}
-(\delta+\sqrt{\delta^2+|V|^2})\\
V^*
\end{array}
\right)_.\label{12}
\end{align}
Therefore, we obtain the following equation for the amplitude of $V$
\begin{align}
|V|={\frac{2|\delta| \frac{|\psi_1|}{|\psi_2|}}{\frac{|\psi_1|^2}{|\psi_2|^2}-1}}_.\label{13}
\end{align}
This means that we can calculate $|V|$ from the ratio of electron densities between the two corners $|\psi_1|^2/|\psi_2|^2$ evaluated from the numerical result of the tight-binding model. We set the parameters $\delta=10^{-6}$, $\gamma=0.2$, and $\lambda=1$ and show our result of $|V|$ between the two corners of the `H'-shaped system, marked with green circles in Fig. 5(a), for various values of $M$ in Fig. 5(d). This figure shows that the amplitude of $|V|$ decay exponentially by a factor of about $10^{-0.6991}\sim 0.2$ with increasing the distance $M$ of the two corners by one unit cell. Equation $(\ref{6})$ obtained in the previous research revealed that the corner states decay with a ratio $-\gamma/\lambda$ along the edge when $\delta=0$\cite{s2}, and we note that it also holds for $\delta \neq 0$. Thus, the above calculation agrees well with this analytical result because $10^{-0.6991}\simeq \gamma/\lambda=0.2$ and we conclude that we can evaluate the penetration depth of the corner states via inteference between them. This is because the penetration depth of the corner states toward the bulk is shorter than that along the edge. Hoppings within other pairs of neighboring corners decay in the similar way.

\section{Conclusion}
In this paper, we reveal characteristics of the penetration depth of the corner states in a higher-order topological insulator. We show that when we change the energy of the corner states toward the end of the edge gap, the penetration depth along the edge diverges toward infinity but that toward the bulk does not. This characteristic is analogous to that of edge states of topological insulators. Meanwhile, corner states have a different aspect from edge states in that corner states have two kinds of penetration depths, that along the edge and that toward the bulk, and they behave differently. We also showed an analytical solution of the corner states when an on-site potential at the corner site is added, and it is expressed in terms of elliptic integrals. The anisotropic and complex behaviors of the penetration depths are encoded in this complex form of the corner-state wavefunction. We also demonstrated that the behavior of the penetration depth along the edge appear as a hybridization between corner states on neighboring corners. Thus, this enables us to experimentally observe the behaviors of the penetration depth.

In this paper, we used the BBH model showing the HOTI phase. Meanwhile, the anisotropic behaviors of penetration depths in this paper are also expected in non-topological corner states in general systems with corner states. Namely, also in nontopological systems, one can physically expect that as corner-state energies approach edge bands or bulk bands, the penetration depth along the edge or into the bulk diverges. Indeed, because the $C_{4}$ symmetry is broken in our calculation, the corner state 
is not topological in a strict sense. Thus the properties of the penetration depths found in this paper do not stem from topological properties but they are expected in a broad class of systems. Furthermore, also in three-dimensional systems, where bulk, surface and hinge bands generally exist, similar behaviors of the corner states are expected. A detailed study on three-dimensional systems is left for a future work.

\begin{acknowledgment}


This work was supported by Japan Society for the Promotion of Science (JSPS) KAKENHI Grants No. JP18H03678, and No. JP20H04633, and by Elements strategy Initiative to Form Core Research Center (TIES), from MEXT Grant Number JP-MXP0112101001.

\end{acknowledgment}

\appendix
\section{Wavefunctions of Edge States in the BBH Model}
In Sec. $2.3$, we discussed edge states of the BBH model in a semi-infinite geometry in Fig. 2(a), which extends from $m=-\infty$ to $m=\infty$, while $n$ takes only positive integer values. In this Appendix, we derive the analytic form of the wavefunctions. The Bloch Hamiltonian at the wavenumber $k$ along the $x$ direction, written in the basis in the lexicographical order $(n,\alpha)=(1,1),(1,2),(1,3),(1,4),(2,1),(2,2),\cdots$, is as follows
\begin{align}
H_{\rm{semi}}(k)=\left(
\begin{array}{ccccc}
A & B & & &\\
B^{\dagger} & A &B & & \\
& B^{\dagger}& A & B & \\
& & B^{\dagger}& \ddots & \\
& & & &
\end{array}
\right)_,
\end{align}
where
\begin{align}
A=\left(
\begin{array}{cccc}
0&0&\gamma+\lambda e^{ik}& \gamma \\
0&0&-\gamma&\gamma+\lambda e^{-ik}\\
\gamma+\lambda e^{-ik}&-\gamma&0&0\\
\gamma&\gamma+\lambda e^{ik}&0&0
\end{array}
\right)_,
\end{align}
\begin{align}
B=\left(
\begin{array}{cccc}
0&0&0&\lambda \\
0&0&0&0\\
0&-\lambda&0&0\\
0&0&0&0
\end{array}
\right)_.
\end{align}
We consider eigenstates of the Bloch Hamiltonian that decay exponentially along the $y$ direction. Namely, we assume the form of the edge-state wavefunction in the following Schr\"{o}dinger's equation
\begin{align}
H\left(
\begin{array}{c}
\bm{\phi} \\
r\bm{\phi} \\
r^2\bm{\phi} \\
\vdots \\
\end{array}
\right)
=E\left(
\begin{array}{c}
\bm{\phi} \\
r\bm{\phi} \\
r^2\bm{\phi} \\
\vdots \\
\end{array}
\right)_,
\end{align}
where $\bm{\phi}$ is a four-component vector and $r$ $(|r|<1)$ describes a decay rate of the edge state into the bulk. From the above equations, we obtain the following equations.
\begin{align}
B^{\dagger}\bm{\phi}&=0,\label{edge1}\\
(A+Br)\bm{\phi}&=E\bm{\phi}.\label{edge2}
\end{align}
From Eq. (\ref{edge1}), $\bm{\phi}$ is given by the following form
\begin{align}
\bm{\phi}=\left(
\begin{array}{c}
0\\
c_1\\
0\\
c_2
\end{array}
\right)_,\label{edge3}
\end{align}
where $c_1$ and $c_2$ are functions of $k$. We substitute Eq. (\ref{edge3}) for Eq. (\ref{edge2}), and we get the following equations
\begin{align}
(\gamma+\lambda r)c_2&=0,\\
(\gamma+\lambda e^{-ik})c_2&=Ec_1,\\
(-\gamma-\lambda r)c_1&=0,\\
(\gamma+\lambda e^{ik})c_1&=Ec_2.
\end{align}
It allows a nontrivial wavefunction only when
\begin{align}
r&=-{\frac{\gamma}{\lambda}}_,\\
E^2&=(\gamma+\lambda e^{-ik})(\gamma+\lambda e^{ik})\notag\\
&=\gamma^2+\lambda^2+2\lambda \gamma \cos{k}.
\end{align}
From these equations, the energy bands of the edge state are given by $E_{\rm{edge}}=\pm \sqrt{\lambda^2+\gamma^2+2\gamma \lambda \cos{k}}$, and they lie in the regions $|\lambda|-|\gamma|\leq|E|\leq|\lambda|+|\gamma|$ when $|\lambda|>|\gamma|$.

\section{Green Function Method for the Analytic Expression of the Corner States}

\subsection{Simple square-lattice model}
We begin with a simple tight-binding model on a square lattice. We consider a spinless tight-binding Hamiltonian 
with one state per each site forming a square lattice. We put this model on a $xy$-plane, with each site
labeled as $(m,n)$, where $m$ and $n$ are integers representing the $x$ and $y$ coordinates, respectively. 
Let $\lambda\ (>0)$ be a hopping amplitude to the nearest neighbor sites
on the square lattice. The model is shown in Fig.~\ref{fig:square}(a). We express the wavefunction in
a form $u_{m,n}$ at the $(m,n)$ site. Then the bulk eigenvalue problem is written as
\begin{align}
Eu_{m,n}=\lambda u_{m,n+1}+\lambda u_{m,n-1}+\lambda u_{m+1,n}+\lambda u_{m-1,n}.
\label{eq:rec}
\end{align}
On the other hand, its bulk eigenenergy is given by $E_{\mathbf{k}}=2\lambda(\cos k_x+\cos k_y)$, where $\mathbf{k}=(k_x,k_y)$ is the
Bloch wavevector and we put the lattice constant to be unity. 

We now consider a corner state of this model in a ``wedge'' geometry (Fig.~\ref{fig:square}(c)),  which consists of the sites with $m\geq 1$ and $n\geq 1$. It forms a 90$^\circ$ corner at $(m,n)=(1,1)$ and infinite in the directions $m\rightarrow \infty$ and $n\rightarrow \infty$. 
We put an on-site potential $\Delta$ only at the corner site $(m,n)=(1,1)$. This on-site potential $\Delta$ makes the form of the corner states 
quite nontrivial, as we see later. 
The eigenvalue equation with the on-site potential $\Delta$ for this wedge geometry is given by 
\begin{align}
&Eu_{m,n}=\lambda u_{m,n+1}+\lambda u_{m,n-1}+\lambda u_{m+1,n}+\lambda u_{m-1,n}\nonumber\\& \hspace{3cm}+\Delta\delta_{m,1}\delta_{n,1}\ (m\geq 1,\ 
n\geq 1), 
\\
&u_{m,0}=0\ (m\geq 1),\ u_{0,n}=0\ (n\geq 1)
\end{align}
In the following, we derive an analytic form of its corner state.

\subsubsection{Edge states}
\begin{figure}
\includegraphics[width=8.5cm]{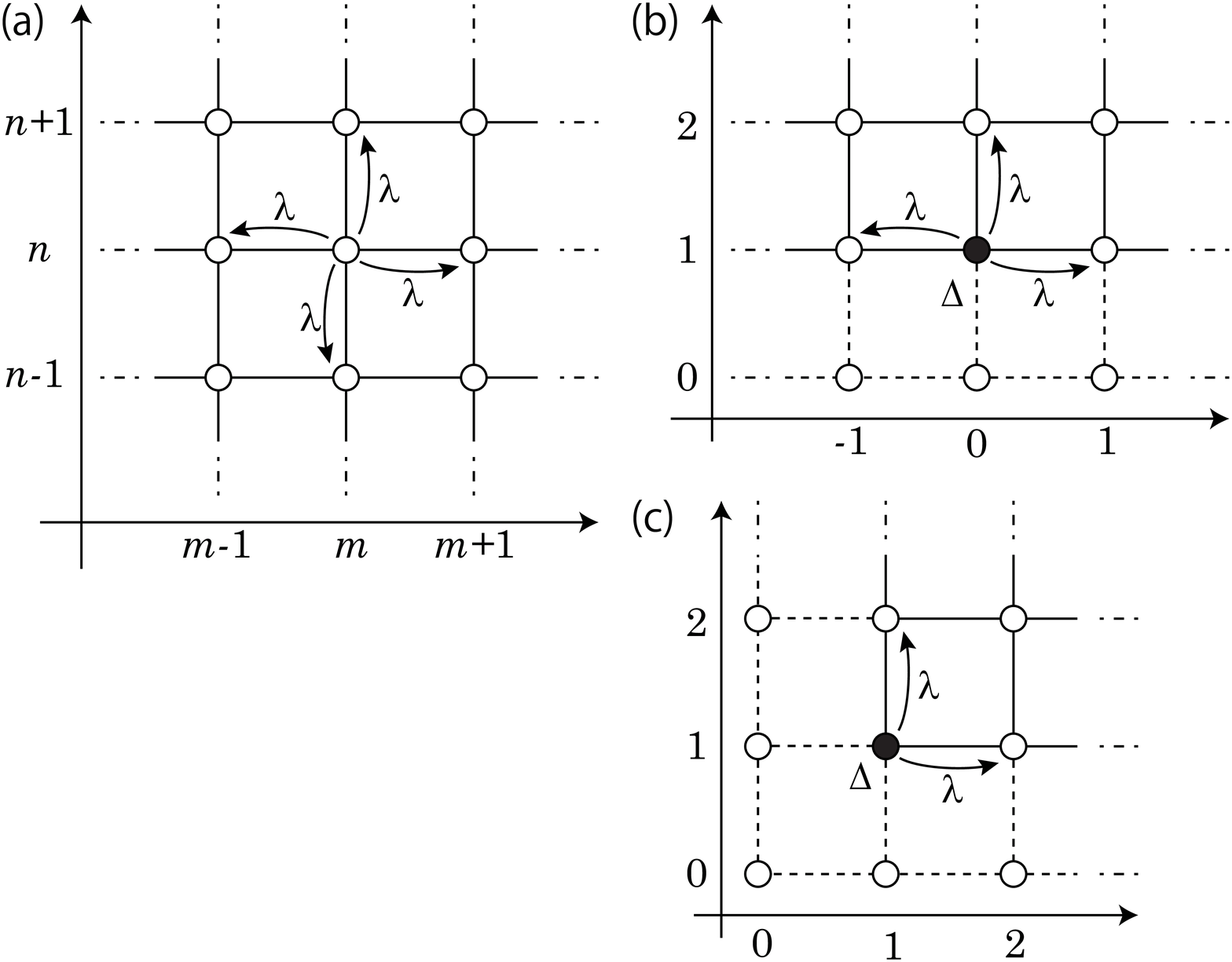}
\caption{(Color online) Tight-binding model on the square lattice. (a) Bulk, (b) semiinfinite, and (c) wedge geometries. We consider the nearest-neighbor hopping $\lambda$ along the solid lines. Later we add
an on-site potential $\Delta$ at (0,1) in (b) and at (1,1) in (c). In the calculation, we add hoppings 
specified by the broken lines in (b) and (c).}
\label{fig:square}
\end{figure}

To solve this problem we first need to solve the eigenvalue problem in a semi-infinite geometry, using the Green's function
developed in Ref. \citen{Cheng}. Here, the semi-infinite geometry refers to the system with sites having $n\geq 1$. 
To this end we 
first consider the system with $n\geq 0$ (see Fig.~\ref{fig:square}(b)), and design the eigenstate to vanish on the line $n=0$.
In this case, 
we rewrite the eigenvalue equation in the following form.
\begin{align}
u_{m,n+1}=\left(-u_{m-1,n}+\frac{E}{\lambda}u_{m,n}-u_{m+1,n}\right)-u_{m,n-1}+p_{m,n}\  (n\geq 1),
\label{eq:umn+1}
\end{align} 
where we put a source term $p_{m,n}$ by hand for later convenience. The case with $p_{m,n}=0$ corresponds to 
the original eigenvalue problem.

Equation (\ref{eq:umn+1}) is a recursion relation with respect to the index $n$, determining the eigenstate in the 
$(n+1)$th row from those in the $n$th row and in the $(n-1)$th row. To write down the eigenstate in a convenient 
way, we introduce a formal series \cite{Cheng}
\begin{align}
u_n(x)\equiv\sum_{m=-\infty}^{\infty}u_{m,n}x^m,
\end{align}
where we introduced a parameter $x$. Note that this is a formal series for the purpose of expressing the values of $u_{m,n}$ in 
a compact way. Thus, we do not discuss convergence/divergence of this series. 
With this formal series, the recursion relation (\ref{eq:umn+1}) is rewritten as 
\begin{align}
&u_{n+1}(x)=\left(-x+\frac{E}{\lambda}-x^{-1}\right) u_n(x)-u_{n-1}(x)+p_n(x)\nonumber \\
 &\hspace{3cm}  (n\geq 1),
\label{eq:un+1}
\end{align} 
where $p_n(x)\equiv\sum_{m=-\infty}^{\infty}p_{m,n}x^m$.
We solve this recursion relation with boundary conditions
\begin{align}
u_0(x)=f(x),\ u_1(x)=g(x),
\end{align}
where $f(x)$ and $g(x)$ are arbitrary functions of $x$, representing the boundary conditions at
$n=0$ and $n=1$ in formal series. 
This can be solved by means of the Green's function \cite{Cheng}.
Let us introduce two Green's functions $G_n(x)$ and $K_n(x)$, where $G_n(x)$ is a
solution for a boundary condition $u_0(x)=1$, $u_1(x)=0$
and $K_n(x)$ is a
solution for a boundary condition $u_0(x)=0$, $u_1(x)=1$. 
Then the solution for Eq.~(\ref{eq:un+1}) is expressed as \cite{Cheng}
\begin{align}
&u_n(x)=f(x)G_n(x)+g(x)K_n(x)\nonumber\\
&\hspace{3cm} +\sum_{j=0}^{n-2}K_{n-1-j}(x)p^{j+1}(x),
\end{align}
Following the method in Ref.~\citen{Cheng}, one can write down the Green's functions
\begin{align}
G_n(x)&=-\frac{1}{\tilde{\tau}}\left\{\ \left(\frac{\tilde{\sigma}+\tilde{\tau}}{2}\right)^{n-1}
-\left(\frac{\tilde{\sigma}-\tilde{\tau}}{2}\right)^{n-1}\right\},
\\K_n(x)&=\frac{1}{\tilde{\tau}}\left\{\ \left(\frac{\tilde{\sigma}+\tilde{\tau}}{2}\right)^{n}
-\left(\frac{\tilde{\sigma}-\tilde{\tau}}{2}\right)^{n}\right\},
\end{align}
where $\tilde{\sigma}=E/\lambda-x-\frac{1}{x}$ and $\tilde{\tau}=\sqrt{\tilde{\sigma}^2-4}$. 

We are interested in eigenstates which vanish at $n=0$. Thus we put $f(x)=0$. Moreover, for the
subsequent calculation of the corner state, we put a source term with $p_{0,1}=C$ and $p_{m,n}=0$ 
($(m,n)\neq (0,1)$) with $C$ being a constant,
leading to $p_1(x)=C$ and $p_n(x)=0$ ($n\neq 1$). 
Then the eigenstate is written as 
\begin{align}
u_n(x)&=g(x)K_n(x)+CK_{n-1}(x)\nonumber\\
&=
\frac{1}{\tilde{\tau}}\left[g(x)
\left\{\ \left(\frac{\tilde{\sigma}+\tilde{\tau}}{2}\right)^{n}
-\left(\frac{\tilde{\sigma}-\tilde{\tau}}{2}\right)^{n}\right\}\right.\nonumber\\
&\ \ \left.+C\left\{\ \left(\frac{\tilde{\sigma}+\tilde{\tau}}{2}\right)^{n-1}
-\left(\frac{\tilde{\sigma}-\tilde{\tau}}{2}\right)^{n-1}\right\}\right].
\label{eq:uedge}
\end{align}
This represents eigenstates in a semi-infinite geometry with the boundary conditions given above. 

We now focus on edge states, which are localized near the edge and decay toward $n\rightarrow \infty$. 
Because 
the bulk band represented by $E=2\lambda(\cos k_x+\cos k_y)$ lies in the region $-4\leq E/\lambda\leq 4$, 
bound states at the edge should lie outside of this region. Such edge states can exist, when there is a strong potential 
near the boundary, as we show in the following. 
Thus, to calculate edge states, we put $E>4\lambda$ here as an example.
The other case with $E<-4\lambda$ can be treated similarly. 

To calculate $u_{m,n}$ from $u_n(x)$, we will calculate a contour integral along the unit circle $|x|=1$ in the complex $x$-plane, 
and so we here consider the case with $x$ being a complex number with $|x|=1$. Then 
it follows that $-2\leq x+x^{-1}\leq 2$, and from the definitions of $\tilde{\sigma}$ we obtain $\tilde{\sigma} >2$. 
We take the branch of $\tilde{\tau}=\sqrt{\tilde{\sigma}^2-4}$ such that $\tilde{\tau}>0$. Then, since
$\frac{\tilde{\sigma}-\tilde{\tau}}{2}\cdot\frac{\tilde{\sigma}+\tilde{\tau}}{2}=1$, we
get $0<\frac{\tilde{\sigma}-\tilde{\tau}}{2}<1<\frac{\tilde{\sigma}+\tilde{\tau}}{2}$. Thus, Eq.~(\ref{eq:uedge}) describes a bound state
only when the term proportional to $(\frac{\tilde{\sigma}+\tilde{\tau}}{2})^n$ vanishes, i.e. when $g(x)=-C\frac{\tilde{\sigma}-\tilde{\tau}}{2}$. 
Under this condition, we get
\begin{align}
u_n(x)
&=
-{C}\left(\frac{\tilde{\sigma}-\tilde{\tau}}{2}\right)^{n}.
\end{align}
We can calculate $u_{m,n}$ by expanding this $u_n(x)$ as a formal power series of $x$, which 
is accomplished via a contour integral
\begin{align}
u_{m,n}=\frac{1}{2\pi i}\oint_{|x|=1} \frac{u_n(x)}{x^{m+1}}dx
&=
-\frac{C}{2\pi i}\oint_{|x|=1} \left(\frac{\tilde{\sigma}-\tilde{\tau}}{2}\right)^{n}\frac{dx}{x^{m+1}}.
\end{align}
It can be also rewritten as
\begin{align}
&u_{m,n}=
-\frac{C}{2\pi }\int_0^{2\pi} d\theta e^{-im\theta}\left(\frac{E}{2\lambda}-\cos\theta-
\sqrt{\left(\frac{E}{2\lambda}-\cos\theta\right)^2-1}\right)^n\nonumber\\
&=
-\frac{C}{2\pi }\int_0^{2\pi} d\theta \cos m\theta \left(\frac{E}{2\lambda}-\cos\theta-
\sqrt{\left(\frac{E}{2\lambda}-\cos\theta\right)^2-1}\right)^n\nonumber \\
&\hspace{4cm} (n\geq 1),
\label{eq:umn}
\end{align}
which is an elliptic integral and cannot be written in terms of elementary functions. 
One can directly check that this solution satisfies the recursion relation (\ref{eq:umn+1}), with a source term $p_{0,1}=C$. In particular, when $n=1$, 
the recursion relation is 
\begin{align}
Eu_{m,1}=\lambda u_{m,2}+\lambda u_{m+1,1}+\lambda u_{m-1,1}-\lambda C\delta_{m,0}.
\end{align}
An important step of our calculation is to 
regard the source term $-\lambda C\delta_{m,0}$ to be an on-site potential $\Delta$ at the 
$(m,n)=(0,1)$ site. This identification is achieved by the following equality:
\begin{align}
&-\lambda C=\Delta \cdot u_{0,1}\nonumber\\\ \ &=
-\frac{C\Delta}{2\pi }\int_0^{2\pi} d\theta \left(\frac{E}{2\lambda}-\cos\theta-
\sqrt{\left(\frac{E}{2\lambda}-\cos\theta\right)^2-1}\right).
\end{align}
Therefore, we get
\begin{align}
\Delta=\frac{2\pi\lambda}{\int_0^{2\pi} d\theta \left(\frac{E}{2\lambda}-\cos\theta-
\sqrt{\left(\frac{E}{2\lambda}-\cos\theta\right)^2-1}\right)}.
\label{eq:U}\end{align}
This is the relationship between the on-site potential $\Delta$ at the (0,1) site  and the energy $E$ of the edge state. If the on-site potential $\Delta$ is given, Eq.~(\ref{eq:U}) determines the eigenenergy $E$ of the 
edge state, and then the eigenstate is given by Eq.~(\ref{eq:umn}).

\subsubsection{Corner state}
\begin{figure}
\includegraphics[width=8.5cm]{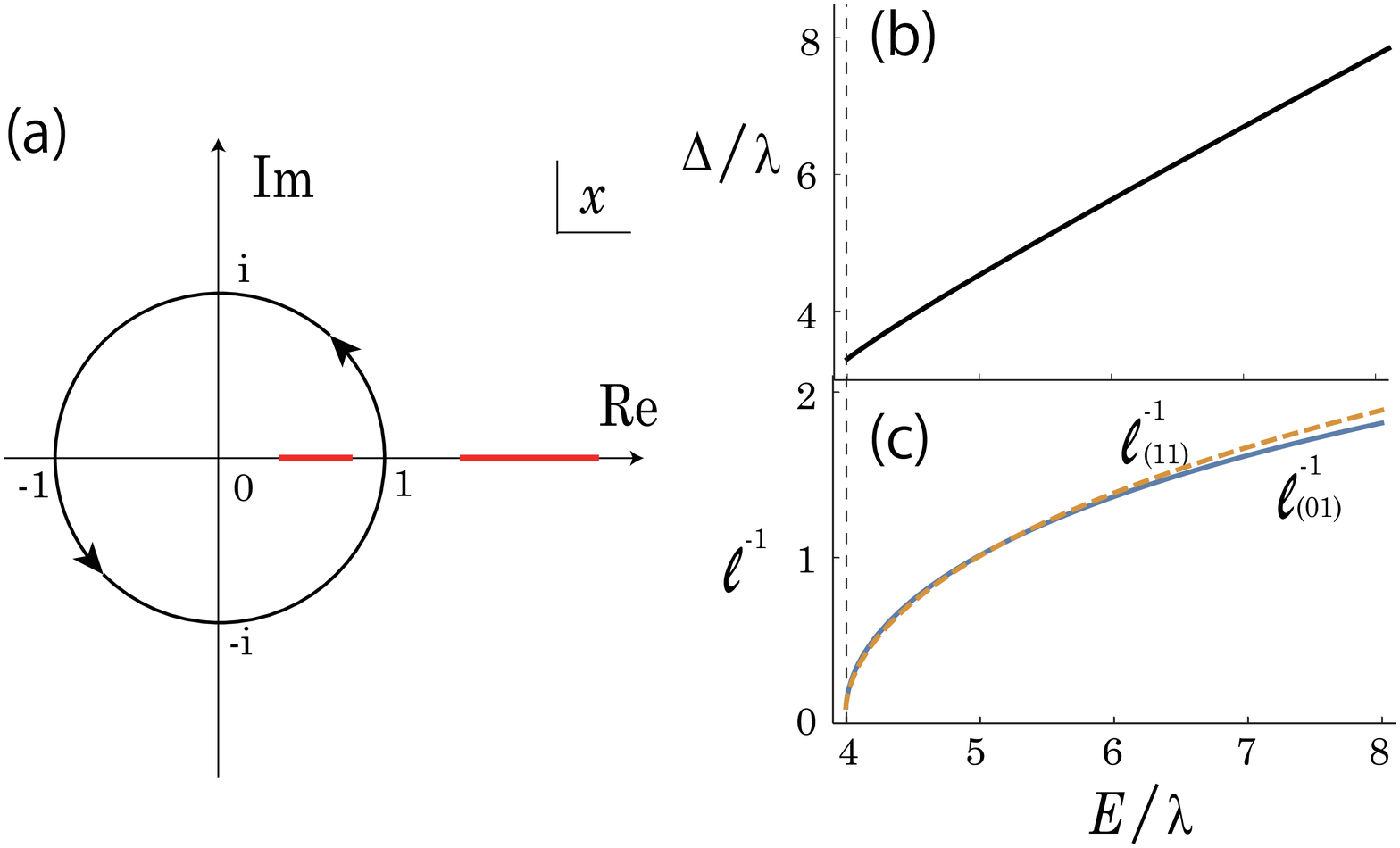}
\caption{(Color online) Corner state of the square-lattice model with on-site potential. (a) Contour and branch cuts  of the integral in Eq.~(\ref{eq:umn-corner}). The branch cuts shown in red have end points specified by $\tilde{\sigma}=\pm 2$. (b) Relationship between the on--site potential $\Delta$ at the (1,1) site and the eigenenergy $E$. (c) Inverses of two localization lengths, $\ell_{(01)}^{-1}=\ln{\left|\tilde{u}_{1,n}/\tilde{u}_{1,n+1}\right|}$ and 
$\ell_{(11)}^{-1}=\frac{1}{\sqrt{2}}\ln{\left|\tilde{u}_{m,m}/\tilde{u}_{m+1,m+1}\right|}$ for a large $m$ in the calculation.}
\label{fig:squareUl}
\end{figure}

Based on the analytic form of the edge state, we can analytically calculate the wavefunction of the 
corner state in the wedge geometry (Fig.~\ref{fig:square}(c)), when an on-site potential exists at the corner site.
We consider the wedge geometry with sites at $(m,n)$ ($m\geq 1$, $n\geq 1$), and an on-site potential $\Delta$ only at the $(1,1)$ site. 
The edge state (\ref{eq:umn}) is a bound state for the semi-infinite geometry, and it decays for  $n\rightarrow \infty$. From this edge-state wavefunction (\ref{eq:umn}) ranging from positive $m$ to negative $m$,
we can directly construct a wavefunction $\tilde{u}_{m,n}$  for the corner state for the wedge geometry, by replacing the source term from $p_{m,n}=C\delta_{m,0}\delta_{n,1}$
to $p_{m,n}=C\delta_{m,1}\delta_{n,1}-C\delta_{m,-1}\delta_{n,1}$, representing an antisymmetric source term under $m\leftrightarrow -m$. 
In terms of the formal series $p(x)$, it amounts to a
replacement 
\begin{align}
p_1(x)=C \ \rightarrow\ p_1(x)=C\left(x-\frac{1}{x}\right).
\end{align}
This antisymmetric source term leads to an antisymmetric superposition of two edge-state wavefunctions, which leads to a
 wavefunction vanishing on the line $m=0$. 
By following the same procedure as we adopted in the calculation of edge states, we get
\begin{align}
\tilde{u}_n(x)
&=
-{C}\left(x-\frac{1}{x}\right)\left(\frac{\tilde{\sigma}-\tilde{\tau}}{2}\right)^{n}.
\end{align}
and 
\begin{align}
\tilde{u}_{m,n}
&=
-\frac{C}{2\pi i}\oint_{|x|=1} \left(x-\frac{1}{x}\right)
\left(\frac{\tilde{\sigma}-\tilde{\tau}}{2}\right)^{n}\frac{dx}{x^{m+1}}.
\label{eq:umn-corner}
\end{align}
It can be also rewritten as 
an integral
\begin{align}
\tilde{u}_{m,n}&=
-\frac{2C}{2\pi }\int_0^{2\pi} d\theta \sin m\theta \sin\theta\nonumber\\&\ \ \cdot\left(\frac{E}{2\lambda}-\cos\theta-
\sqrt{\left(\frac{E}{2\lambda}-\cos\theta\right)^2-1}\right)^n\ (n\geq 1).
\label{eq:umn-corner2}
\end{align}
In this case, the on-site potential $\Delta$ at the (1,1)-site is related to the eigenenergy $E$ by $-\lambda C=\Delta\cdot\tilde{u}_{1,1}$, i.e.
\begin{align}
\Delta=\frac{\pi\lambda}{\int_0^{2\pi} d\theta \sin^2\theta \left(\frac{E}{2\lambda}-\cos\theta-
\sqrt{\left(\frac{E}{2\lambda}-\cos\theta\right)^2-1}\right)}.
\label{eq:delta-energy}
\end{align}
Equation (\ref{eq:umn-corner}) is a contour integral along the unit circle. The integrand has  two 
branch cuts, one inside and the other outside of the unit circle (see Fig.~\ref{fig:squareUl}(a)). This integral, when rewritten as
Eq.~(\ref{eq:umn-corner2}), is an elliptic integral and cannot be expressed in terms of elementary
functions. We show the relationship between the on-site potential $\Delta$ and the eigenenergy $E$ obtained from Eq. (\ref{eq:delta-energy}) in Fig. \ref{fig:squareUl}(b).

We note that this system in the wedge geometry preserves mirror symmetry $M$
with respect to the line $m=n$, and the resulting corner state is expected to possess this symmetry.
Indeed, the eigenstate is symmetric under the mirror symmetry $M$, i.e. $\tilde{u}_{m,n}=\tilde{u}_{n,m}$. This symmetry is not at all
obvious in Eq.~(\ref{eq:umn-corner}), but it can be 
directly shown by using a transformation of integation variables $E/\lambda-x-x^{-1}=\tilde{x}+\tilde{x}^{-1}$. By this transformation, 
the
interior and the exterior of the contour are exchanged while $m$ and $n$ are exchanged in the integrand, leading to $\tilde{u}_{m,n}=\tilde{u}_{n,m}$. 

We here discuss a penetration depth of corner states. The corner states decay away from the corner, 
and how rapidly they decay into the bulk may depend on the directions. One can define its penetration depth along the $(0\ 1)$ direction (i.e. along the edge) is given by $\tilde{u}_{1,n}\propto e^{-n/\ell_{(01)}}$, 
leading to $e^{-1/\ell_{(01)}}=\left|\tilde{u}_{1,n+1}/\tilde{u}_{1,n}\right|$ i.e. $1/\ell_{(01)}=\ln{\left|\tilde{u}_{1,n}/\tilde{u}_{1,n+1}\right|}$ ($n$: large). 
On the other hand, we can also consider its decay along the (1\ 1) direction parallel to the $m=n$ line. 
Along the (1\ 1) direction, we consider another penetration depth $\ell_{(11)}$ into the bulk along the $m=n$ direction
by 
 $1/\ell_{(11)}=\frac{1}{\sqrt{2}}\ln{\left|\tilde{u}_{m,m}/\tilde{u}_{{m+1},{m+1}}\right|}$ ($m$: large). 
In Fig.~\ref{fig:squareUl}(c), we plot the two penetration depths $\ell_{(01)}$ and $\ell_{(11)}$.
We observe that these two penetration depths are different but quite close to each other. We attribute this isotropic behavior to the fact that this 
lattice Hamiltonian is well approximated by an isotropic contimuum Hamiltonian $H=\frac{-\lambda}{2}\nabla^2$ near the band edge. Thus the 
distribution is circular around the corner site, and to see this clearly we added a factor $\frac{1}{\sqrt{2}}$ in the definition of $\ell_{(11)}$.

\subsection{Corner state of the BBH model}
We can use the similar trick to calculate the corner state for the BBH model in a wedge geometry, when
an on-site potential is added on the corner site: $(m,n)=(1,1)$ and $\alpha=2$. 
Nonetheless, we cannot follow the same procedure, because it is not straightforward to construct a corner state out of edge states, unlike the previous simple tight-binding model. 
Nonetheless, one can guess what is the wavefunction, and  can prove that it is the exact eigenstate of the corner state. 

We again write the wavefunction $u_{m,n}^{(\alpha)}$ in terms of a formal series in terms of $x$:
\begin{align}
u_{n}^{(\alpha)}(x)&=\sum_{m=-\infty}^{\infty}
u_{m,n}^{(\alpha)}x^m.
\end{align}
Similar to the square-lattice model, we assume a solution in a
form
$\bm{u}_{n}(x)={}^t(u_n^{(1)},u_n^{(2)},u_n^{(3)},u_n^{(4)})=\oint_{|x|=1}Q(x)\frac{R^n}{x^{m+1}}{\bm{v}}dx$, with a parameter $R$. Then 
we obtain
\begin{align}
\left(\begin{array}{cccc}
&&\gamma+\frac{\lambda}{x} &\gamma+\lambda R\\
&&-\gamma-\frac{\lambda}{R}&\gamma+\lambda x\\
\gamma+\lambda x  &- \gamma-\lambda R&& x\\
\gamma+\frac{\lambda}{R}& \gamma+\frac{\lambda}{x} &&\\
\end{array}\right){\bm{v}}
=E{\bm{v}}.
\end{align}
This leads to
\begin{align}
&\lambda\gamma\left(R+\frac{1}{R}+x+\frac{1}{x}\right)+2\lambda^2+2\gamma^2=E^2,\\
&\bm{v}=a_1\bm{v}_1+a_2\bm{v}_2,\ 
{\bm{v}}_1=\left(\begin{array}{c}
0\\E\\-\gamma-\lambda R\\ \gamma+\frac{\lambda}{x}
\end{array}
\right),\ 
{\bm{v}}_2=\left(\begin{array}{c}
E\\0\\ \gamma+\lambda x\\ \gamma+\frac{\lambda}{R}
\end{array}
\right),\end{align}
with constants $a_j$.
This equation for $R$ gives two solutions:
\begin{align}
& R=\frac{\sigma\pm\tau}{2}, \\ 
&\sigma=\frac{E^2-2\lambda^2-2\gamma^2}{\lambda\gamma}-x-\frac{1}{x}, \ \ 
\tau=\sqrt{\sigma^2-4}
\end{align}
In this paper, we focus on the case $0<\gamma<\lambda$. First we consider the case where the corner state inside the edge gap, i.e. $-(\lambda-\gamma)<E<\lambda-\gamma$.
Similar to the calculation on the square-lattice model stated earlier, we express the solution as a
contour integral along the unit circle $|x|=1$. We take the branch of $\tau=\sqrt{\sigma^2-4}$ such that $\tau>0$ when $|x|=1$. Thus, by assuming $|x|=1$ and the energy being within 
the gap, the two solutions of $r$ satisfies
$\frac{\sigma-\tau}{2}<-1<\frac{\sigma+\tau}{2}<0$. Thus in the corner state, only the 
solutions proportional to $\left(\frac{\sigma+\tau}{2}\right)^n$ is allowed, and those proportional to 
 $\left(\frac{\sigma-\tau}{2}\right)^n$ should be absent because it diverges at $n\rightarrow \infty$. Thus we conjecture a solution of the form
\begin{align}
\bm{u}_{m,n}&=\oint_{|x|=1}h(x)\frac{1}{x^{m+1}}\left(\frac{\sigma+\tau}{2}\right)^n\left(
\begin{array}{c}
0\\E\\-\gamma-\lambda\frac{\sigma+\tau}{2}\\ \gamma+\frac{\lambda}{x}
\end{array}
\right)\frac{dx}{2\pi i},\nonumber\\
&\hspace{3cm} (m\geq1, n\geq 1),
\label{eq:umn-BBH}
\end{align}
 where $h(x)$ is a function of $x$ to be determined. Here, we put ${\bm{v}}={\bm{v}}_1$, because 
 we found that the first component vanishes from numerical calculations of corner states. We will show that 
 this conjecture holds.

In order to derive the function $h(x)$, we impose a boundary condition for the corner state
\begin{align}
u_{m,0}^{(3)}=0\ (m\geq 2), 
\label{eq:um0}
\end{align}
in order that this represents an eigenstate in the wedge geometry. 
In our eigenvector (\ref{eq:umn-BBH}), the integrand contains a branch cut inside the 
unit circle, which  makes the integral nonzero in general. Thus, to guarantee the condition
Eq.~(\ref{eq:um0}), we can assume the form of the function $h(x)$ to be
\begin{align}
&h(x)=\frac{H(x)}{\gamma+\lambda\frac{\sigma+\tau}{2}}
\end{align}
and $H(x)$ is a rational function of $x$. With this assumption, $u_{m,0}^{(3)}$ has no
branch cut inside the unit circle.  
Furthermore, we also impose
 \begin{align}
u_{0,n}^{(4)}=0\ (n\geq 2).
\label{eq:un0}
\end{align}
To guarantee this condition, we put $H(x)=P(x)(x-\frac{1}{x})/(\gamma+\frac{\lambda}{x})$, 
which makes the integrand in the expression of $u_{0,n}^{(4)}$ an odd function of $\theta$ where $x=e^{i\theta}$, 
and thus the integral over $\theta$ vanishes. 
We cannot uniquely determine $H(x)$ only from these conditions, but as a trial, we put
$P(x)=1$ and $H(x)=(x-\frac{1}{x})/(\gamma+\frac{\lambda}{x})$, Therefore, we now have the
following form for the eigenvector:
\begin{align}
\bm{u}_{m,n}&=\oint_{|x|=1}\frac{x-\frac{1}{x}}{\left(\gamma+\frac{\lambda}{x}\right)
\left(\gamma+\lambda\frac{\sigma+\tau}{2}\right)}\nonumber\\
&\ \ \cdot\frac{1}{x^{m+1}}\left(\frac{\sigma+\tau}{2}\right)^n\left(
\begin{array}{c}
0\\E\\-\gamma-\lambda\frac{\sigma+\tau}{2}\\ \gamma+\frac{\lambda}{x}
\end{array}
\right)\frac{dx}{2\pi i}\nonumber\\&\hspace{3cm}  (m\geq1, n\geq 1).
\label{eq:umn-BBH2}
\end{align}

Equation (\ref{eq:umn-BBH2}) is our conjecture so far, but it does not agree with numerical calculations
of corner states. 
In order to improve the conjecture, we focus on the following symmetry property.
The BBH model with the on-site potential in the wedge geometry preserves 
mirror symmetry $M$ with respect to the $m=n$ line. This is represented by a matrix
\begin{align}
\hat{M}=\left(\begin{array}{cccc}1&&&\\&1&&\\&&&-1\\&&-1&\end{array}\right),
\end{align}
and it leads to the symmetry property of the corner state
\begin{align}
&u_{m,n}^{(1)}=u_{n,m}^{(1)}, \ 
u_{m,n}^{(2)}=u_{n,m}^{(2)}, \label{eq:symm-u}\\
&u_{m,n}^{(3)}=-u_{n,m}^{(4)}, \ 
u_{m,n}^{(4)}=-u_{n,m}^{(3)}.
\label{eq:symm-u2}
\end{align}
Let us check whether Eq.~(\ref{eq:umn-BBH2}) satisfies these symmmetry 
conditions. We use the duality relation similar to the square-lattice model, and 
we find that Eq.~(\ref{eq:umn-BBH2}) itself does not satisfy the symmetry conditions
(\ref{eq:symm-u}) (\ref{eq:symm-u2}). Meanwhile, one can show that by changing the integral contour from the unit circle to the
loop $C$ depicted in Fig.~4(a), we show that the wavefunction
\begin{align}
\bm{u}_{m,n}&=\oint_{C}\frac{x-\frac{1}{x}}{\left(\gamma+\frac{\lambda}{x}\right)
\left(\gamma+\lambda\frac{\sigma+\tau}{2}\right)}\nonumber\\
&\ \ \cdot\frac{1}{x^{m+1}}\left(\frac{\sigma+\tau}{2}\right)^n\left(
\begin{array}{c}
0\\E\\-\gamma-\lambda\frac{\sigma+\tau}{2}\\ \gamma+\frac{\lambda}{x}
\end{array}
\right)\frac{dx}{2\pi i}\  (m\geq1, n\geq 1)
\label{eq:umn-BBH3}
\end{align}
satisfies the symmetry conditions (\ref{eq:symm-u}) (\ref{eq:symm-u2}). In this case, one branch cut and one pole are inside the contour $C$, and another branch cut and another pole are outside. Through a duality transformation
\begin{align}
& \tilde{x}+\frac{1}{\tilde{x}}=\frac{E^2-2\lambda^2-2\gamma^2}{\lambda\gamma}-x-\frac{1}{x}
\end{align}
from $x$ to $\tilde{x}$, the inside and outside of the contour $C$ are exchanged, and the symmetry conditions (\ref{eq:symm-u})  (\ref{eq:symm-u2}) are
shown. In fact, this result (\ref{eq:umn-BBH3}) perfectly matches with our result of numerical diagonalization, 
and we thus establish that 
(\ref{eq:umn-BBH3}) is the analytic form of the corner state wavefunction. 
The on-site potential $\Delta$ is obtained similarly to the square lattice model stated earlier. We get
\begin{align}
\Delta=\frac{\lambda}{\gamma u^{(2)}_{1,1}}.
\end{align}

In a numerical calculation, this integral is evaluated as a
sum of an integral along the unit circle and a contribution coming from the 
pole 2 at $x=-\frac{\lambda}{\gamma}$. When $-(\lambda-\gamma)<E<\lambda-\gamma$, it is written as
\begin{align}
\bm{u}_{m,n}&=\int_{0}^{2\pi}d\theta \frac{ie^{-i m\theta}\sin{\theta}}{\pi \left(\gamma+\lambda e^{-i\theta}\right)
\left(\gamma+\lambda\frac{\sigma(\theta)+\tau(\theta)}{2}\right)}\nonumber\\
&\ \ \cdot \left(\frac{\sigma(\theta)+\tau(\theta)}{2}\right)^n\left(
\begin{array}{c}
0\\E\\-\gamma-\lambda\frac{\sigma(\theta)+\tau(\theta)}{2}\\ \gamma+\lambda e^{-i\theta}
\end{array}
\right)\ \nonumber \\
& + \frac{-\gamma^2+\lambda^2}{\gamma^3}
\left(\gamma+\lambda \frac{\bar{\sigma}+\bar{\tau}}{2}\right)^{-1}\nonumber\\
&\ \ \cdot\left(-\frac{\gamma}{\lambda}\right)^{m+1}\left(\frac{\bar{\sigma}+\bar{\tau}}{2}\right)^n\left(
\begin{array}{c}
0\\E\\-\gamma-\lambda\frac{\bar{\sigma}+\bar{\tau}}{2}\\ 0
\end{array}
\right)\  (m\geq1, n\geq 1)
\label{eq:umn-BBH4}
\end{align}
where
\begin{align}
&\sigma(\theta)=\frac{E^2-2\lambda^2-2\gamma^2}{\lambda\gamma}-2\cos\theta,\ \ \tau(\theta)=\sqrt{\sigma(\theta)^2-4}\ (>0),
\\
&\bar{\sigma}=\frac{E^2-\lambda^2-\gamma^2}{\lambda\gamma}, \ \ 
\bar{\tau}=\sqrt{\bar{\sigma}^2-4}\ (>0).
\end{align}

Next we consider the case where the corner state lies between the edge band and the neighboring bulk band, 
i.e. $-\sqrt{2}(\lambda-\gamma)<E<-(\lambda+\gamma)$ or
 $(\lambda+\gamma)<E<\sqrt{2}(\lambda-\gamma)$.
 In this case, the relative positions of the branch cuts and the poles change, and this leads to the change of
 the sign of $\sqrt{\sigma^2-4}$ in the residue at the pole 2 at $x=-\frac{\lambda}{\gamma}$, which amounts to 
 an extra minus sign in front of $\bar{\tau}$:
 \begin{align}
\bm{u}_{m,n}&=\int_{0}^{2\pi}d\theta \frac{ie^{-i m\theta}\sin{\theta}}{\pi \left(\gamma+\lambda e^{-i\theta}\right)
\left(\gamma+\lambda\frac{\sigma(\theta)+\tau(\theta)}{2}\right)}\nonumber\\
&\ \ \cdot \left(\frac{\sigma(\theta)+\tau(\theta)}{2}\right)^n\left(
\begin{array}{c}
0\\E\\-\gamma-\lambda\frac{\sigma(\theta)+\tau(\theta)}{2}\\ \gamma+\lambda e^{-i\theta}
\end{array}
\right)\ \nonumber \\
& +\frac{-\gamma^2+\lambda^2}{\gamma^3}
\left(\gamma+\lambda\frac{\bar{\sigma}-\bar{\tau}}{2}\right)^{-1}\nonumber\\
&\ \ \cdot\left(-\frac{\gamma}{\lambda}\right)^{m+1}\left(\frac{\bar{\sigma}-\bar{\tau}}{2}\right)^n\left(
\begin{array}{c}
0\\E\\-\gamma-\lambda\frac{\bar{\sigma}-\bar{\tau}}{2}\\ 0
\end{array}
\right)\  (m\geq1, n\geq 1).
\label{eq:umn-BBH5}
\end{align}
These formulas are used in the calculation of the corner-state wavefunctions in the main text.

\end{document}